\documentclass[useAMS]{mn2e}
\usepackage{graphicx}
\title[The Evolution of type 1 AGN in the IR
   (15$\mu$m)]{The Evolution of type 1 AGN in the IR
   (15$\mu$m). \\ The view from ELAIS-S1 
\thanks{Based on observations with the Infrared Space
   Observatory ({\it ISO}). {\it ISO} is an ESA project 
   with instruments funded by
   ESA Member States (especially the PI countries: France, Germany,
   The Netherlands and the United Kingdom) and with the participation
   of ISAS and NASA. }}
\author[I. Matute et al.]
       {I. Matute,$^1$ F. La Franca,$^1$ F. Pozzi,$^{2,3,4}$ 
	C. Gruppioni,$^{4,5}$ C. Lari,$^2$ G. Zamorani,$^4$
	\newauthor
	D.M. Alexander,$^6$ L. Danese,$^7$ S. Oliver,$^8$ S. Serjeant $^9$ 
	and M. Rowan-Robinson$^9$\\
        $^1$Dipartimento di Fisica, Universit\`a ``Roma Tre'', via della Vasca
	     Navale 84, I-00146 Roma, Italy\\
	$^2$Istituto di Radioastronomia del CNR, via Gobetti 101, 
		I-40129 Bologna, Italy\\
	$^3$Dipartimento di Astronomia, Universit\`a di Bologna, 
         viale Berti Pichat 6, I-40127 Bologna, Italy \\
	$^4$Osservatorio Astronomico di Bologna, via Ranzani 1, 
		I-40127 Bologna, Italy\\
	$^5$Osservatorio Astronomico di Padova, vicolo dell$'$Osservatorio 5,
	 I-35122 Padova, Italy\\
	$^6$Department of Astronomy \& Astrophysics, 525 Davey Laboratory, 
		Penn State University, University Park, PA 16802, USA\\
	$^7$SISSA, via Beirut 4, I-34014 Trieste, Italy\\
	$^8$Astronomy Centre, CPES, University of Sussex, Falmer, 
		Brighton, BN1 9QJ, UK\\
	$^9$Imperial College of Science, Technology and Medicine, 
		Prince Consort Road, London SW7 2BZ. UK}
\date{Accepted 2002 February ??.
      Received 2002 February ??;
      in original form 2001 December 8}

\pagerange{\pageref{firstpage}--\pageref{lastpage}}
\pubyear{2002}

\begin{document}

\maketitle

\label{firstpage}

\begin{abstract}

   We present the 15$\mu$m luminosity function of AGN1 (QSO + Seyfert
   1). Our sample of 21 high-redshift sources is selected from the
   Preliminary Analysis catalogue in the S1 field of the European
   Large Area {\it ISO} Survey (ELAIS). To study the cosmological
   evolution of the AGN1 luminosity function, our sample has been
   combined with a local sample of 41 sources observed by {\it IRAS}.
   We find that the luminosity function of AGN1 at 15$\mu$m is fairly
   well represented by a double-power-law-function. There is evidence
   for significant cosmological evolution consistent with a pure
   luminosity evolution (PLE) model $L(z)$$\propto$$(1+z)^{k_L}$, with
   $k_{L}=3.0-3.3$.  The value of $k_{L}$ depends on the existence or
   not of an evolution cut-off at redshift $\sim$2, and on the adopted
   cosmology.  From the luminosity function and its evolution we
   estimate a contribution of AGN1 to the Cosmic Infrared Background
   (CIRB) of $\nu I_{\nu} \sim 6 \times 10^{-11}$W m$^{-2}$ sr$^{-1}$
   at 15$\mu$m. This corresponds to $\sim 2-3\%$ of the total observed
   CIRB at this wavelength.  Under the usual assumptions of unified
   models for AGN, the expected contribution of the whole AGN
   population to the CIRB at 15$\mu$m is $10-15\%$.
\end{abstract}


\section{Introduction}
The measurement of the luminosity function (LF) of Active Galactic Nuclei
(AGN) and its evolution provides fundamental information on the
accretion history of the Universe, the physics related to it,
how structures have been formed and on the contribution of this
population to the Cosmic Background.

Historically, AGN have been  classified using their optical
characteristics and  divided into two categories: type 1 (AGN1)
and 2 (AGN2), according to the presence or absence of broad emission lines
in their optical spectra (we will use this definition throughout this
paper).

Unified models assume a dusty, IR emitting, torus around the central
engine. These models distinguish AGN1 and AGN2 depending on the torus
orientation; AGN1 are galaxies with low absorption from the mid-IR to
the soft X-ray while AGN2 are objects with the largest column
densities (e.g. Antonucci 1993; Urry \& Padovani 1995).

The strength  of the AGN1 LF evolution has been determined in the optical,
soft and hard X-rays (e.g. Boyle et al. 2000; Miyaji et al. 2000; La
Franca et al. 2002), while in the mid infrared (5-30 $\mu$m: mid-IR)
it is still unknown. The only measure available at these wavelengths, for
AGN1, is the local LF at 12 $\mu$m from {\it IRAS} data (Rush, Malkan \&
Spinoglio, 1993; RMS hereafter).

The situation for AGN2 is quite different because they are strongly absorbed
and thus difficult to select and/or spectroscopically identify.
Statistically significant samples of AGN2 need to be built up in the far
infrared or the hard X-rays where absorption is less significant. However,
in these cases the optical spectroscopic identification is also a 
difficult task.

For these reasons the contribution of AGN to the Cosmic Infrared
Background (CIRB) is still unknown. The only estimate is based 
on  AGN1 evolution measured at other wavelengths, under
assumptions on the AGN spectral energy distribution (SED), and
the AGN1/AGN2 ratio (e.g. Granato, Danese \& Franceschini 1997).

We have spectroscopically identified 25 AGN1 at 15$\mu$m from the
Preliminary Analysis (PA) catalogue of the European Large Area {\it
ISO} Survey (ELAIS). Our spectroscopic sample is the only
one able to probe the high redshift population of AGN1 in this
waveband. Other  published {\it ISO} samples (i.e. CFRS, Flores
et al. 1999; ISO-HDF north, Aussel et al. 1999; ISO-HDF south, Oliver
et al. 2002) do not have enough sources and/or spectroscopic
identifications.

Here we present our measurement of the evolution of LF of AGN1 at 15$\mu$m
using our data combined with the 12$\mu$m local sample from RMS. The
data are presented in section 2.  In section 3 we describe the method
used to derive the LF, while the results are presented and discussed 
in section 4.

\section[]{The Sample}
ELAIS is the largest single open time project conducted by {\it ISO} (Oliver
et al. 2000), mapping an area of 12 deg$^2$ at 15$\mu$m with ISOCAM
and at 90$\mu $m with ISOPHOT.  Four main fields were chosen in order
to reduce the effects of cosmic variance (N1, N2, N3 in the north
hemisphere and S1 in the south) at high Ecliptic latitudes
($|\beta|>40^{\scriptscriptstyle{o}}$) on selected areas of low
extinction.

A Preliminary Analysis (PA) catalogue was produced for the 4 main
fields (Serjeant et al. 2000). The Final Analysis (FA) 
uses the {\em Lari technique} and the catalogue 
of the S1 field 
(J2000,$\, \alpha: 00^{\scriptscriptstyle{h}}34^
{\scriptscriptstyle{m}}44.4^{\scriptscriptstyle{s}}\, ,\delta:
{-43\degr28\arcmin12\arcsec}$, covering an area of 3.96 deg$^2$) has
been recently released by Lari et al. (2001).

Optical identifications were possible due to an extensive R-band
CCD survey, performed with the ESO/Danish 1.5m and the ESO/MPE 2.2m
telescopes.  The spectroscopic follow-up program was carried out with
the AAT at the Anglo-Australian Observatory (AAO), the 3.6m and NTT at
\mbox{ESO/La Silla}.
The selection of targets for spectroscopic identification 
was carried out randomly in order to
cover uniformly $\sim$70\%  of the sources in the magnitude 
range $17$$<$R$<$$20$. The dominant
class is star-forming galaxies (about 50\%), however
AGN (type 1 + 2) constitute a significant fraction of the 
identifications in this magnitude range ($\sim$30\%). 
The ELAIS-S1 field has also been
completely covered in the radio at 1.4 GHz down to 0.3 mJy (Gruppioni
et al. 1999), and in  50$\%$ of its area in the X-rays with {\it BeppoSAX}
(Alexander et al. 2001).

In order to assess our sample of optically identified PA sources, we
took advantage of the already completed FA catalogue in the ELAIS-S1
region (Lari et al. 2001).  The FA catalogue contains 462 sources at
5$\sigma$ confidence level, while the PA catalogue we used contains
762 sources, as it includes also less significant sources.

All the PA sources with spectroscopic identifications were analyzed,
and the flux recomputed with the {\it Lari Technique} at $>4\sigma$
confidence level: 23 out of 25 PA AGN1 were confirmed and used for our
statistical analysis. 

The fluxes of the PA catalogue were recomputed according to the FA
calibration by using all the objects in common, plus the PA objects
with existing spectroscopy and detected also with the {\it Lari
Technique} at $> 4\sigma$ confidence level.  The effective area, as a
function of flux, of our PA selected sample has been statistically
estimated by comparing the PA ``corrected'' counts (i.e. including
only those detected at $> 4\sigma$ confidence level with the {\it Lari
Technique}) with the counts derived from the FA catalogue by Gruppioni
et al. (2002).

Our final sample consists of 21 AGN1 (0.3$<$$z$$<$3.2, see Figure 1)
with $f_{15\mu m}$$>$$1\mathrm{mJy}$ and $17$$<$R$<$$20$ (two objects 
fainter than R=20 fall outside these limits). 

To represent the local mid-infrared population of type 1 AGN
we have combined our data with a sample of 41 AGN1 belonging to the
RMS sample selected from the IRAS {\it Faint Source Catalog, Version
2} (Moshir et al. 1991). This sample is complete down to 300 mJy at
12$\mu$m.

\begin{figure}
\centering
\includegraphics[angle=90,width=8cm]{{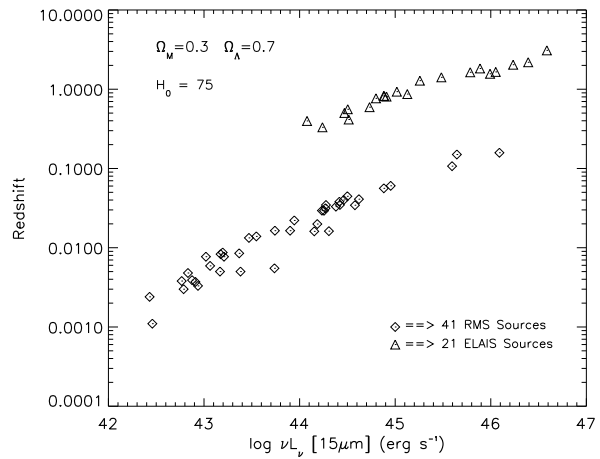}}
\caption{Distribution on the Luminosity-redshift plane of 
	the type 1 AGN sources used in computing 
	the luminosity function.}
\label{FigGam}%
\end{figure}

\section{The Luminosity Function}
In this work we define $L_{15}$ as $\nu L_{\nu}$ at 15$\mu$m.
Detailed $k$-corrections (over the 15$\mu$m LW3 filter from {\it ISO}
and the 12$\mu$m filter from {\it IRAS}) are necessary in order to
derive the 15$\mu$m luminosities for our sources, as well as to
convert the 12$\mu$m luminosity from the RMS sample into a 15$\mu$m
luminosity. We adopted the mean spectral energy distribution (SED) of
radio-quiet QSOs from Elvis et al. (1994) as representative of our
AGN1 population. The standard $k$-correction was computed following
Lang (1980). Figure 1 illustrates the luminosity-redshift space distribution of
all AGN1 used in this study.

A parametric, unbinned maximum likelihood method was used to fit
simultaneously the evolution and luminosity function parameters
(Marshall et al. 1983) of the combined sample.  Since ELAIS
identifications were carried out in the R-band magnitude interval
17.0$<$R$<$20.0, a factor $\mathbf{\Theta}(z,L)$, which takes into
account the optical limits, was introduced into the likelihood
function.  $\mathbf{\Theta}(z,L)$ represents the probability that a
source with a given luminosity at 15$\mu$m ($L_{15}$) and redshift
$z$, had a R-magnitude within the limits of the sample, and was
derived taking into account the 1$\sigma$ internal spread in the
assumed SED (Elvis et al. 1994). The R-band $k$-corrections were taken
from Natali et al. (1998). In order to show the dependence of our
sample completeness on the assumed SED we plot the expected relation
(and its 1$\sigma$ spread) between the IR and optical fluxes (Figure
2). From this we estimate that our sample of AGN1 is about 76\%
complete. The missing fraction of objects is expected to be found in
the interval 20.0$<$R$<$22.0.  The function to be minimized can
therefore be written as $\mathrm{S}=-2\,\ln\,\cal L$, being $\cal L$
the likelihood function,
\begin{center}
\begin{displaymath}
\mathrm{S}=
-2\sum_{i=1}^{\mathrm{N}} \ln[\Phi(z_{i},\!L_{i})]  + 2 \!\!\int\!\!\!\!\!\int 
\!\Phi(z,\!L) \Omega(z,\!L)  \mathbf{\Theta}(z,\!L)  \frac {\textstyle dV}
{\textstyle dz}dzdL
\end{displaymath}
\end{center}

\noindent where N is the total number of sources in the two samples,
$\Omega(z,L)$ is the fractional area of the sky over which a
source with luminosity $L_{15}$ and redshift $z$ could have been
observed, $(dV/dz)$ is the differential comoving volume and
$\Phi(z,L)$ the space and luminosity density of the sources being considered. 

   \begin{figure} 
	\centering
   	\includegraphics[angle=90,width=8cm]{{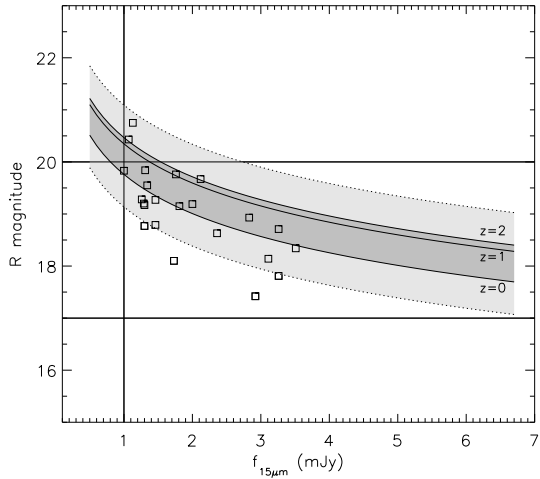}} \caption{The
   apparent R-magnitudes as a function of 15$\mu$m flux, according to
   the SED from Elvis et al. (1994) and R-band $k$-corrections from
   Natali et al. (1998). Continuous lines represent the central values
   at redshifts $z$=0, 1 and 2. Dashed lines represent the maximum
   spread at 1$\sigma$ confidence level. Vertical and horizontal lines
   are the IR and optical limits of the ELAIS-S1 survey.  Plotted {\it
   squares} correspond to our AGN1. } 
	\label{FigTheta} 
	\end{figure}

   \begin{figure} \centering
   \includegraphics[angle=90,width=8.5cm]{{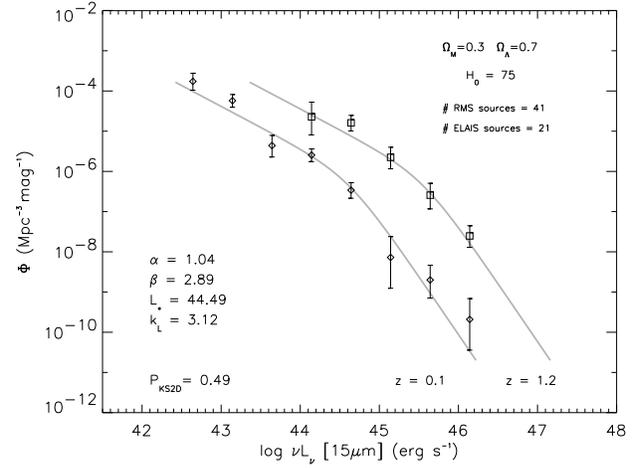}}
   \caption{Best PLE luminosity function fit (see also table 1).  Two
   redshift intervals are shown.  The points correspond to the space
   densities of the observed sources, corrected for evolution within
   the redshifts intervals. Sources with $z$=[0.0,0.2] are represented
   by {\it diamonds} and correspond to the RMS sources. Sources with
   z=[0.2,2.2] are represented by {\it squares}, and represent the
   ELAIS-S1 population.  Lines plotted are the mid-LF at the central
   redshift of the interval considered.  The error bars are based on
   Poisson statistics at the 68$\%$ confidence level.  }
   \label{FigGam} \end{figure}

Following studies of AGN1 in the optical and in the X-ray, we
adopted a smooth double power-law for the space density distribution
in the local Universe,

$$ \frac{d\Phi({L}_{\rm 15},z=0)}{d{\rm log}{L}_{\rm 15}} =
\frac{\Phi^{*}}{(L_{\rm 15}/{{L}^{*}_{\rm
15}})^{\alpha} + (L_{\rm 15}/{{L}^{*}_{\rm
15}})^{\beta}}$$
and tested a Pure Luminosity Evolution (PLE) model of the form:
$$ L_{\rm 15}(z) = L_{\rm 15}(0)(1+z)^{k_{L}}. $$

Confidence regions for each parameter have been obtained by computing
$\Delta \mathrm{S}(\equiv
\Delta \chi^2)$ at a number of values around the best-fitting 
($\mathrm{S}_{min}$) parameter, while allowing 
  the other parameters to float (see Lampton, Margon \&
Bowyer 1976). The 68\% confidence levels quoted correspond to $\Delta
 \mathrm{S = 1.0}$. The normalization factor $\Phi^{*}$ was determined
by requiring agreement with the observed total number of sources (ELAIS + RMS).

Bidimensional Kolmogorov-Smirnov (2D-KS) probabilities (see Peacock 1983;
Fasano \& Franceschini 1987) were derived for each of the best fitting PLE
models.

We computed the LF both in a
($\Omega_{\mathrm{m}}$,$\Omega_\Lambda$)=(1.0,0.0) and
($\Omega_{\mathrm{m}}$,$\Omega_\Lambda$)=(0.3,0.7) cosmology,
assuming $H_0=75$ $\mathrm{km} \,\mathrm{s}^{-1}\,\mathrm{Mpc}^{-1}$.

\linespread{1.0}
\begin{table*}{}
\caption[]{Parameter values of the fit of the Luminosity Function}
	\begin{center}
	\begin{tabular}{ccccccccc}
	\hline

& $\alpha$ & $\beta$ & ${\mathrm{log} \, L_*}^a$ & $k_{L}$ & $z_{cut}$ & 
${\mathrm{log \,\Phi^{*}}}^b$ & $\mathrm{CIRB}^{c}$ & $P_{2DKS}$\\

\hline\hline
$(\Omega_{\mathrm{m}},\Omega_{\Lambda})=(1.0,0.0) $ &
$1.12_{-0.07}^{+0.07}$ & $2.93_{-0.30}^{+0.31}$ &
$44.51_{-0.06}^{+0.06}$ & ${2.98}_{-0.22}^{+0.16}$ & $----$ &
$-5.67$ & $5.15$ & $0.28$\\
\\
$(\Omega_{\mathrm{m}},\Omega_{\Lambda})=(1.0,0.0) $ &
$1.13_{-0.07}^{+0.07}$ & $2.90_{-0.26}^{+0.20}$ &
$44.49_{-0.04}^{+0.07}$ & ${3.30}_{-0.18}^{+0.15}$ &
$2.0(\mathrm{fixed})$ & $-5.65$ & $6.41$ & $0.16$\\
\\
$(\Omega_{\mathrm{m}},\Omega_{\Lambda})=(0.3,0.7)$ &
$1.04_{-0.08}^{+0.07}$ & $2.89_{-0.26}^{+0.32}$ &
$44.49_{-0.05}^{+0.06}$ & ${3.12}_{-0.20}^{+0.13}$ & $----$ &
$-5.67$ & $6.70$ & $0.49$\\
\\
$(\Omega_{\mathrm{m}},\Omega_{\Lambda})=(0.3,0.7)$ &
$1.03_{-0.05}^{+0.10}$ & $2.89_{-0.26}^{+0.29}$ &
$44.49_{-0.08}^{+0.05}$ & ${3.30}_{-0.19}^{+0.15}$ &
$2.0(\mathrm{fixed})$ & $-5.66$ & $5.66$ & $0.54$\\

	\hline \end{tabular}
	\end{center} \label{FitRes}
        \begin{list}{}{}
   	\item[ ] Errors correspond to 68\% confidence level for a single 
		interesting parameter. 
	\item[$a.$] $L_*$ corresponds to $\nu
	L_{\nu}$ and is given in $\mathrm{erg \,s}^{-1}$.  
	\item[$b.$] The normalization,
	$\mathrm{log}\,\Phi^{*}$, is given in $\mathrm{Mpc}^{-3}\,
	\mathrm{log}\,L ^{-1}$. 
	\item[$c.$] Contribution ($ \nu I_{\nu}$)
	of type 1 AGN to the 15$\mu$m Mid-IR background in 
	units of $10^{-11} 
	{\mathrm{W m^{-2} sr^{-1}}}$.
		
	\end{list}
\end{table*}
\linespread{1.0}

\section{Results and Discussion}
Figure 3 and Table 1 show the results of the fit for the 2 cosmologies
considered together with the 2D-KS probabilities. The points plotted
in Figure 3 correspond to the space densities of the observed sources,
corrected for evolution within the redshift intervals as explained in
La Franca \& Cristiani (1997).

Our ($\Omega_{\rm m}$,$\Omega_\Lambda$)=(1.0,0.0) local LF is not fully
consistent with the estimate of RMS. Our bright slope $\beta \sim 2.9$
is steeper than the estimate $\beta \sim 2.1$ of RMS. This difference
can probably be  attributed both to a steeper high-$z$ LF and to the
effects of evolution on the brightest bins used by
RMS.

The Mid-Infrared LF, for either cosmologies, is steeper than
that observed in the optical and in the soft and hard X-ray. The faint
15$\mu$m slope is $\alpha\sim 1.1$ while the bright slope is
$\beta\sim 2.9$. In the optical Boyle et al. (2000) found $\alpha\sim
0.5$ and $\beta\sim 2.5$ (here we quote slope-1 in order to convert to
$d\Phi/d\log\,L$ units). In the soft and hard X-rays the
faint slope is $\alpha\sim 0.7$ while the bright slope is $\beta\sim
2.0$ (Miyaji, Hasinger \& Schmidt 2000; La Franca et al. 2002).

According to the blue optical spectra of our identified AGN1, we can
exclude the possibility that a different population has been
selected. As a consequence this steepening at longer wavelength of the
LF, taken at face value, would imply a dependence of the spectral
energy distribution of AGN1 on luminosity and/or redshift. For
example, a larger IR contribution from the SED of the fainter AGN1
(possibly caused by a more relevant contribution from the hosting
galaxy) could be the origin of the observed behaviour. A more detailed
discussion on this issue will be carried out when the results from the
identifications of the FA catalogue from ELAIS are available.

Similarly to what has been observed in the optical, the evolution at 15$\mu$m
is fairly well represented by a PLE model with $k=3.0-3.3$. Our data
are not statistically significant enough to probe the presence of a cut-off in
the evolution ($z_{cut}$) as observed in the optical and X-ray
bands. As shown in Table 1, our fits are consistent with a 
typical value $z_{cut}=2.0$, similar to what has been found from optical
and X-ray studies (e.g. Boyle et al. 2000; Miyaji et al. 2000). All
the fits show a good 2D-KS probability, with larger values
in a $(\Omega_{\mathrm{m}},\Omega_{\Lambda})=(0.3,0.7)$
Universe.

Our estimates of the 15$\mu$m LF allow us to compute the contribution
of AGN1 to the 15$\mu$m CIRB.  Direct measurements of the CIRB in the
5-100$\mu$m range are difficult. Hauser et al. 2001, report an upper
limit of 470 and 500 $\mathrm{nW m^{-2} sr^{-1}}$ at 12 and 25$\mu$m
respectively, based on the attenuation of the $\gamma$-ray photons
(see also Hauser et al. 1998). On the other hand, a lower limit of
$3.3\pm 1.3$ $\mathrm{nW m^{-2} sr^{-1}}$ at 15$\mu$m is found from
the integrated light of source counts derived from observations with
ISOCAM (Altieri et al. 1999).

The intensity of the Cosmic Background at 15$\mu$m is given by,
$$\mathrm{I_\nu}=\frac{1}{4\pi}\int{dL_{15}\int{dz \,\frac{dV}{dz}\,
f_\nu(L_{15},z) \,\Phi(L_{15},z)}} $$ where $f_\nu(L_{15},z)$
represents the observed flux density of a source with an intrinsic
luminosity $L_{15}$ at redshift $z$.  According to our derived PLE
models, we estimate a contribution of the AGN1 to the CIRB at 15$\mu$m
(in units of $\nu I_{\nu}$) in the range $5.2-6.7
\times 10^{-11}\;\mathrm{W m^{-2} sr^{-1}}$, depending on the cosmology adopted
and on the existence or not of a redshift cut-off ($z_{cut}$, see
Table 1). The integral has been computed for $\mathrm{log}\,
L_{15}(z=0)>42$ up to $z=5.0$. These values correspond to about
\mbox{$2-3\%$} of the integrated light measured by Altieri et
al. (1999).  If we assume that the ratio of AGN2 to AGN1, which is
about 4 locally (e.g. Lawrence 1991; Maiolino \& Rieke, 1995), does
not change substantially with redshift and that the shape and
evolution of the luminosity function of AGN2 is similar to those of
AGN1, the total contribution for AGN to the background measured by
Altieri et al. (1999) would be $\sim10-15\%$. This value is in rough
agreement with what is predicted by the models of dusty tori in AGN
from Granato et al. (1997).  The predictions from Xu et al. (2001) on
the contribution of AGN to the CIRB at 15$\mu$m is about a factor of 2
lower than our estimate. This difference is probably due to two
reasons. First, the slope of their local LF of AGN1+AGN2 is flatter
than our estimates at faint luminosities. Second, their assumed
evolution (a PLE with $k_L$=3.5 up to $z$=1.5 and then dropping as
$k_L$=-3) has a lower $z_{cut}$ and a faster decay at high redshift.

\section*{Acknowledgments}
We acknowledge J. Kotilainen, A. Verma and
M. Linden-Vernle for carrying out part of the CCD R-band photometric
observations and data reduction. Based on observations collected at
the European Southern Observatory, Chile.  IM acknowledges the receipt
of a fellowship within the CRA-OAP-IAC agreement. This research has
been partly supported by ASI contracts ARS-99-75, MURST grants
Cofin-98-02-32, Cofin-0002036 and EC TMR Network programme (FMRX-CT96-0068).

\bsp

\label{lastpage}

\end{document}